\documentclass[11pt,a4paper]{article}
\pdfoutput=1

\usepackage{tikz}
\usetikzlibrary{arrows,shapes,through}

\usepackage{amsmath}
\usepackage{amssymb,array,amsfonts,amsthm}

\usepackage[utf8]{inputenc}
\usepackage[OT1]{fontenc}

\usepackage{verbatim}
\usepackage{booktabs}
\usepackage{url}

\usepackage{floatflt}
\usepackage[algoruled,linesnumbered]{algorithm2e}
\usepackage{marginnote}

\begin{document}
\title{Pure Parsimony Xor Haplotyping}
\author{%
Paola~Bonizzoni%
\thanks{Paola~Bonizzoni is with DISCo, Univ.~Milano-Bicocca},
Gianluca~Della~Vedova%
\thanks{Gianluca~Della~Vedova is with Dip.~Statistica, Univ.~Milano-Bicocca},
Riccardo~Dondi%
\thanks{Riccardo~Dondi is with Dip.~Scienze dei Linguaggi, della Comunicazione e degli Studi Culturali, Univ.~Bergamo},
Yuri~Pirola%
\thanks{Yuri~Pirola is with DISCo, Univ.~Milano-Bicocca},
and~Romeo~Rizzi%
\thanks{Romeo~Rizzi is with DIMI, Univ.~Udine}%
}
\date{Preliminary version}

\maketitle

\begin{abstract}
The haplotype resolution from xor-genotype data has been recently
formulated as a new model for genetic studies~\cite{Barzuza04Xor}.
The xor-genotype data is a cheaply obtainable type of data
distinguishing  heterozygous from homozygous sites without
identifying the homozygous alleles.
In this paper we propose a formulation  based on a
well-known model used in haplotype inference: pure parsimony.
We exhibit exact solutions of the problem by  providing
polynomial time algorithms for some restricted cases and a
fixed-parameter algorithm for the general case. These results are
based on some interesting combinatorial properties of  a graph
representation of the  solutions.
Furthermore, we show that the
problem has a polynomial time $k$-approximation, where $k$ is the
maximum number of xor-genotypes containing a given SNP.
Finally, we propose a heuristic and produce an
experimental analysis showing that it scales to real-world large
instances taken from the HapMap project.
%, and to even larger synthetic instances.
% Le istanze sintetiche sono molto + piccole di quelle di HapMap
\end{abstract}

\newtheorem{Theorem}{Theorem}[section]
\newtheorem{Lemma}[Theorem]{Lemma}
\newtheorem{Proposition}[Theorem]{Proposition}
\newtheorem{Property}[Theorem]{Property}
\newtheorem{Observation}[Theorem]{Observation}
\newtheorem{Corollary}[Theorem]{Corollary}
\newtheorem{Claim}[Theorem]{Claim}
\newtheorem{Remark}[Theorem]{Remark}

\theoremstyle{definition}
\newtheorem{Definition}{Definition}[section]
\newtheorem{Problem}{Problem}
\newtheorem{Example}{Example}[section]
\newtheorem{Fact}{Fact}[section]
\newtheorem{Rule}{Rule}

\newcommand{\nota}[1]{%
  {\sffamily\bfseries #1}%
  \marginpar{\framebox{$\mathbf{\Leftarrow}$}}%
}
\newcommand{\tyb}{\emph{type b }}
\newcommand{\tya}{\emph{type a }}
\newcommand{\grafo}{\mathcal{G}}
\newcommand{\gadgetgrafo}{\mathcal{VG}}
\providecommand{\card}[1]{\lvert#1\rvert}

\newcommand{\notaestesa}[2]{%
  {\sffamily {\bfseries #1} #2}%
  \marginnote{\centering \Huge \bfseries !}%
}

\newcommand{\solG}{\ensuremath{\mathcal{G}}}

\renewcommand{\emptyset}{\varnothing}

\section{Introduction}
\label{sec:intro}

In this paper we investigate a computational  problem arising in
genetic studies of diploid organisms. In such organisms (which
include all vertebrates), all chromosomes are in two copies, one
inherited from the mother and  one from the father.
Since chromosomes   are almost identical except for specific gene
variants   called \emph{Single Nucleotide Polymorphisms} (or SNPs), changes
between variants are represented by a sequence
of sites, each one bearing a specific value called
\emph{allele}. In almost all cases, for each site at most two
different alleles are present in the population, one of which is
called major and the other one minor. The sequence of alleles
along a chromosome is called \emph{haplotype}, while a
\emph{genotype} is a sequence of unordered pairs of alleles that
appear in each site of the two copies of the chromosome. Haplotype data are
crucial in genetic population studies. The current technology for
finding the two haplotypes of an individual is too expensive to be
used in genetic studies of a population. Fortunately it is much
cheaper to determine the \emph{xor-genotype}  for each individual,
that is  the set of sites for which the individual is
\emph{heterozygous}, i.e. bearing   both a major and a minor allele.
The term xor-genotype derives from the fact that a site is
reported in the genotype  if and only if the two alleles in the
site are different.
Thus a  xor-genotype lists only heterozygous
sites,  while excluding  sites bearing identical alleles, called
\emph{homozygous} sites.

 The problem of reconstructing
the haplotypes resolving a given set of xor-genotypes naturally
arises and represents an interesting case of the process of
inferring haplotypes from general genotypes (\emph{phasing}).

Polynomial time algorithms for the problem have been
developed~\cite{Barzuza04Xor,Barzuza08Xor} in the framework of the
Perfect Phylogeny model, originally proposed by Gusfield~\cite{Gusfield2002}
to solve the phasing problem.
% Rimosso perche' duplicato con il precedente.  (Yuri)
% Computational models for the problem has been recently
% developed~\cite{Barzuza04Xor,Barzuza08Xor} in the framework of the
% Perfect Phylogeny model, originally proposed by
% Gusfield~\cite{Gusfield2002} to solve the phasing problem, solving the
% problem with a reduction to the graph realization problem.

In this paper, we investigate the problem under the  parsimonious
principle that asks for a
smallest set of haplotypes resolving all input xor-genotypes: such
problem will be called  \textsc{Pure Parsimony Xor Haplotyping}
(PPXH).

Let $\Sigma$ be a set of sites (also called characters).
Then a \emph{xor-genotype} (or simply a \emph{genotype}) $x$ is a
non-empty subset of $\Sigma$, and a \emph{haplotype} $h$ is a (possibly
empty) subset of $\Sigma$.
Given two distinct haplotypes $h_1$, $h_2$,  then the pair $(h_1, h_2)$
\emph{resolves} the xor-genotype $x$ iff  $x=h_1 \oplus h_2$,
where $\oplus$ is defined as the classical symmetric difference of $h_1$
and $h_2$, i.e.~the set of characters that are present in exactly one
of $h_1$ and $h_2$.
A set  $H$ of haplotypes resolves a set $X$ of xor-genotypes  if
for each genotype $x \in X$, there exists a pair of haplotypes in $H$
that resolves $x$.

We are now able to formally introduce the problem that we will study in this
paper.

\begin{Problem} \textsc{Pure Parsimony Xor Haplotyping (PPXH).}
The instance of the problem is  a set $X$ of xor-genotypes, and the goal is to
compute a smallest  set $H$ of haplotypes resolving  $X$.
\end{Problem}

The pure parsimony model has been used
%investigated
as an approach to the phasing problem over regular genotypes
(i.e.~where alleles of each homozygous site are
specified)~\cite{Gusfield2003}.
There is a rich literature in this area; in particular the
APX-hardness~\cite{Lancia2004} of the problem and the lack of good
approximation guarantees have led many researchers to the design
of methods based on linear programming techniques to find solutions of
the problem~\cite{BHAR}.
Indeed, the best known approximation algorithm yields
approximation guarantees of $2^{k-1}$ where $k$  is the maximum
number of heterozygous sites appearing in each
genotype~\cite{Lancia2004}.
Restricted cases of the problem with polynomial time solutions
have been studied~\cite{IerselKKS08,Lancia2006}.

In the paper we investigate the PPXH problem mainly by devising
exact solutions of the problem by either considering
fixed-parameter tractability  or polynomial time algorithms for some
restricted instances of the problem.
We introduce a new graph representation of
xor-genotypes and haplotypes, called \emph{xor-graph}, that is crucial
in the study of the PPXH problem.
Indeed most of the results that we will present rely on combinatorial
properties of xor-graphs.

Initially we will show that the PPXH problem is equivalent to the
problem of building a xor-graph with the fewest possible vertices.
Afterwards we design two polynomial time solutions for
restricted instances of the PPXH problem.
Subsequently we
%investigate the fixed-parameter complexity of the
%problem and
design a fixed-parameter algorithm of $O(mn + 2^{k^2}km)$ time complexity, for
$k$ the size of the optimum solution.
Moreover we provide a
$l$-approximation  algorithm, where $l$ is the maximum number of
occurrences  of a character in the set of input genotypes.
Finally we propose a heuristic for the general problem and an
experimental analysis on real and artificial datasets.
The experimental analysis shows that the heuristic is effective on a
large class of instances of various sizes where other methods, such as
the ILP formulation proposed by Brown and Harrower~\cite{BHAR},
are not applicable.
% Affermazione troppo forte/anticipata?

\section{Basic Properties}
\label{sec:xorgraph}

A fundamental idea used in our paper is a graph representation of a feasible
solution. More precisely, given a set $X$ of
xor-genotypes, the
 representation of  a set $H$ of haplotypes resolving $X$ is the
graph $\solG=(H, E)$,  called \emph{xor-graph}  associated with $H$,
where edges of $\solG$ are labeled
by a bijective function
$\lambda: E \to X$ such that, for each edge $e = (h_i, h_j)$,
$\lambda(e) = h_i \oplus h_j$.
% edge $e \in E$ is univocally labeled  by a xor-genotype $\lambda(e)\in X$,
% where $\lambda(e)= h_i\oplus h_j$ for  each edge $e=(h_i,h_j) \in E$.
The labeling $\lambda$ is generalized
to a set $S$ by defining $\lambda(S)=\{\lambda(s) \mid s\in S\}$.
We call \emph{optimal xor-graph for $X$}, a xor-graph associated with an optimal
solution for  $X$ (that is a xor-graph with the minimum number of
vertices).

In this section we state some basic combinatorial properties   of
xor-graphs  that will be used to prove the main results of the
paper.
Among all possible haplotypes, we identify a distinguished haplotype, called
\emph{null} haplotype and
denoted by $h_0$, which corresponds to the empty set.
Since the operation $\oplus$ is associative and commutative, by a
slight abuse of language, given a family $F=\{s_1,\ldots,s_n\} $ of
subsets of a set $\Sigma$ we denote by $\oplus(F)$ the expression $s_1
\oplus s_2 \oplus \cdots \oplus s_n$.
% Let $x$ be a xor-genotype, then we denote by $h(x)$ the haplotype such that
% $h(x) \oplus h_0=x$. In this case the pair $( h(x),h_0)$ is called
% \emph{canonical resolution} of $x$ and $h(x)$ is the canonical haplotype
% of $x$.

The  cycles of a xor-graph satisfy the following property.

\begin{Lemma}
\label{lem:0-sum}
Let $X$ be a set of xor-genotypes, let $\solG$ be a xor-graph
associated with a set of haplotypes resolving   $X$ and let $C$ be
the edge set of a cycle of $\solG$. Then $\oplus(\lambda(C))$
is equal to the empty set.
\end{Lemma}

\begin{proof}
By definition of cycle,  $C$ consists of a set  $\{(h_1,h_2),
(h_2,h_3), \ldots, (h_n,h_{n+1})\}$, with $h_1=h_{n+1}$. By
definition of xor-graph, $\oplus(\lambda(C)) = \oplus_{i=1}^n
\left( h_i
  \oplus h_{i+1} \right) $. By the associativity and commutativity of $\oplus$,
$\oplus(\lambda(C)) = (h_1 \oplus h_{n+1}) \oplus \left( h_2
\oplus
  h_2 \right) \oplus \ldots \oplus \left( h_n \oplus h_n \right) $.
Since $h_1 = h_{n+1}$  and $h\oplus h=\emptyset$ for each $h$,
we obtain $\oplus(\lambda(C))=\emptyset$.
\end{proof}

The above property of cycles of a xor-graph is   sufficient to construct
a set of haplotypes resolving a set of genotypes from a xor-graph.
Let $X$ be an instance of PPXH and let
 $\solG=(V,E)$ be a graph  whose edges  are biunivocally labeled by a
function $\lambda:E \to X$  such that $\oplus(\lambda(C))=
\emptyset$ for each cycle $C$ of the graph.
Then it is immediate to compute a
feasible solution $H$ from $\solG$ where $|H|\le |V|$.
More precisely, we associate a haplotype with each vertex
of $\solG$ as follows.
Associate the null haplotype $h_0$ with any vertex in each  connected
component of $\solG$.
Perform a depth-first visit of each connected component of $\solG$, starting
from the vertex associated with $h_0$.
When visiting a new vertex $v$ of $\solG$ there must exist an edge
$e=(v,w)$ so that the haplotype $w_h$ has been previously assigned to
$w$.
Then associate the haplotype $w_h\oplus \lambda(e)$ with $v$.

It is not hard to verify that our construction guarantee that $H$ is actually a
feasible solution of $X$, that is for each edge $e=(v,w)$ of $\solG$,
$v_h\oplus w_h=\lambda(e)$, where $v_h$ and $w_h$ are respectively the
haplotypes associated with $v$ and $w$. It is trivial to notice that the
property holds for all edges that are part of the spanning forest computed by
the depth-first search. Therefore we can restrict our attention to edges $e=(v,w)$
that are not in such spanning forest. Since $v$ and $w$ are in the same
connected component of $\solG$ the spanning tree $T$ of the connected component
contains both $v$ and $w$. Let $x$ be the least common ancestor of $v$ and $w$
in $T$. By construction the two paths of $T$, both starting from $x$ and
ending one in $v$ and the other in $w$ are edge disjoint. Let us denote by
$P_v$ and $P_w$ respectively the edges of the paths ending in $v$ and
$w$, and let $x_h$ be the haplotype associated with $x$. Now we want to prove
that  $v_h\oplus w_h= \lambda(e)$.
It is immediate to verify
that $v_h=\bigoplus(\lambda(P_v)) \oplus x_h$ and
$w_h=\bigoplus(\lambda(P_w))  \oplus x_h$.
Since the edges in $P_v\cup P_w\cup \{e\}$ form a
simple cycle of $\solG$, by Lemma~\ref{lem:0-sum} we can conclude
$\lambda(e)=\bigoplus(\lambda(P_v)) \oplus \bigoplus(\lambda(P_w))$,
completing the proof.

The following results justify our  attention to connected xor-graphs
and their cuts.

\begin{Lemma}
\label{lem:cut-symbol}
Let $X$ be a set of xor-genotypes and let
$\solG$ be a xor-graph associated with a set $H$ of haplotypes
resolving $X$.  Let $\alpha$ be any character of $\Sigma$. Then the
set $A$ of edges of $\solG$ whose label contains $\alpha$ is a cut
of $\solG$.
\end{Lemma}

\begin{proof}
Let $H_\alpha$ be the subset of $H$ containing the character $\alpha$, and
let $\bar{H}_\alpha=H \setminus H_\alpha$. Let $E'$ be the edges of $\solG$ with an
endpoint in $H_\alpha$ and one in $\bar{H}_\alpha$ (clearly $E'$ is a cut of
$\solG$.) Notice that  $E'$ is exactly the set of edges connecting a haplotype
containing $\alpha$ and a haplotype not containing $\alpha$,
therefore $E'=A$.
\end{proof}

\begin{Lemma}
\label{lem:opt-connected}

Let $X$ be a set of xor-genotypes, and
let  $\solG=(H,E)$ be a disconnected xor-graph for $X$.
Then $\solG$ is not an optimal xor-graph of $X$
\end{Lemma}
\begin{proof}
Since  $\solG$ has at least two  connected components  $C_1$ and $C_2$, we
denote with  $a_1$, $a_2$ two vertices
from $C_1$ and $C_2$ respectively. Construct the set $H'$ from $H$
by replacing each haplotype $h\in C_1$ by $h\oplus a_1$ and  each
haplotype $h\in C_2$  by $h\oplus a_2$.
Since $C_1$ and $C_2$ are not connected, the set of genotypes resolved
by $H'$ is equal to that of $H$.

But both $a_1$ and $a_2$ are replaced by the null haplotype in
$H'$, therefore $|H'|$ is strictly smaller than $|H|$.
\end{proof}

Instances and solutions of the PPXH problem can be represented by
binary matrices.
More precisely, we can have a \emph{genotype matrix} associated with
a set of xor-genotypes and a \emph{haplotype matrix} associated with
a set of haplotypes. In both matrices each column is uniquely
identified by a character in $\Sigma$, while the rows of a
genotype matrix (respectively haplotype matrix) correspond to the
genotypes (resp. haplotypes).

\begin{figure}[htb!]
  \centering
\begin{tikzpicture}[label distance=2mm , scale=.8,every node/.style={inner sep=0pt, minimum width=2pt}]
  \coordinate (x)   at (3,5);
  \coordinate (y) at (0,0);
  \coordinate (z)      at (6,0);
  \node (h1) [label=90:$\emptyset$] at (0,0)    {};
  \node (h5) [label=210:$ad$, pin distance=2cm] at (canvas polar cs:angle=210,radius=3cm) {};
  \node (h2) [label=30:$cd$] at (canvas polar cs:angle=30,radius=3cm) {};
  \node (h6) [label=150:$d$]  at (canvas polar cs:angle=150,radius=3cm) {};
  \node (h4) [label=150:$ac$] at (canvas polar cs:angle=270,radius=3cm) {};
  \node (h3) [label=90:$abcd$] at (canvas polar cs:angle=90,radius=3cm) {};
  \node (h7) [label=330:$e$] at (canvas polar cs:angle=330,radius=3cm) {};

  \foreach \v in {h1, h2, h3, h4, h5, h6, h7}{
    \fill (\v) circle(2.5pt);
  }

  \draw  (h2) -- (h3) node[midway,above right]{$ab$};
  \draw  (h3) -- (h6) node[midway,above left]{$abc$};
  \draw  (h3) -- (h5) node[midway,above left]{$bc$};
  \draw  (h4) -- (h7) node[midway,below right]{$ace$};
  \draw  (h2) -- (h7) node[midway,right]{$cde$};
  \draw  (h1) -- (h7) node[midway,above right]{$e$};
  \draw  (h5) -- (h6) node[midway,left]{$a$};

  \end{tikzpicture}
  \caption{Xor-graph representing the set of haplotypes in Table~\ref{tab:example}.}
  \label{fig:example-xor-graph}
\end{figure}
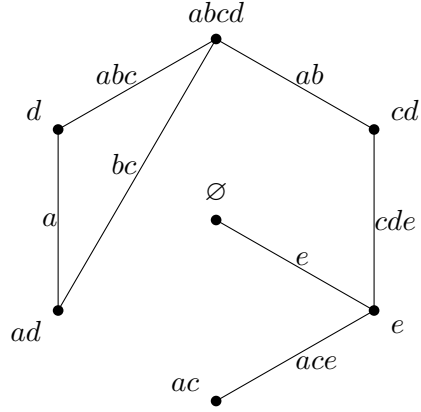

For example let $\Sigma$ be the set $\{a, b, c, d, e\}$ and let $X$ be
the set of xor-genotypes $\{ \{a, b\}, \{a, b, c\},  \{b, c\},  \{c, d,
e\},  \{a\}, \{e\}, \{a, c, e\}\}$.
A possible, albeit suboptimal, set of haplotypes resolving $X$ is
$\{ \emptyset, \{c, d\},  \{a, b, c, d\},  \{a, c\},  \{a, d\},
\{d\}, \{e\}\}$.
The matricial representation of both sets is in Table~\ref{tab:example},
while the associated xor-graph  is represented in
Figure~\ref{fig:example-xor-graph}.

\begin{table}[htb!]
  \caption{Example of genotype (left) and haplotype (right) matrices.}
  \label{tab:example}
  \centering
  \begin{minipage}[t]{4.2cm}
  \centering
    \begin{tabular}{l | l  l l l l}
            & $a$  & $b$  & $c$  & $d$  & $e$ \\\hline
      \rule{0pt}{2.6ex}%
      $g_1$ & $1$  & $1$  & $0$  & $0$  & $0$ \\
      $g_2$ & $1$  & $1$  & $1$  & $0$  & $0$ \\
      $g_3$ & $0$  & $1$  & $1$  & $0$  & $0$ \\
      $g_4$ & $0$  & $0$  & $1$  & $1$  & $1$ \\
      $g_5$ & $1$  & $0$  & $0$  & $0$  & $0$ \\
      $g_6$ & $0$  & $0$  & $0$  & $0$  & $1$ \\
      $g_7$ & $1$  & $0$  & $1$  & $0$  & $1$ \\
    \end{tabular}
  \end{minipage} \hfill
  \begin{minipage}[t]{4.2cm}
  \centering
    \begin{tabular}{l|lllll}
            & $a$  & $b$  & $c$  & $d$  & $e$ \\\hline
      \rule{0pt}{2.6ex}%
      $h_1$ & $0$  & $0$  & $0$  & $0$  & $0$ \\
      $h_2$ & $0$  & $0$  & $1$  & $1$  & $0$ \\
      $h_3$ & $1$  & $1$  & $1$  & $1$  & $0$ \\
      $h_4$ & $1$  & $0$  & $1$  & $0$  & $0$ \\
      $h_5$ & $1$  & $0$  & $0$  & $1$  & $0$ \\
      $h_6$ & $0$  & $0$  & $0$  & $1$  & $0$ \\
      $h_7$ & $0$  & $0$  & $0$  & $0$  & $1$ \\
    \end{tabular}
  \end{minipage}
\end{table}

Given  an ordering of the character
set (that is $\Sigma=\langle\sigma_1, \ldots ,
\sigma_{|\Sigma|}\rangle$), the entry in the $i$-th row and $j$-th
column of a genotype matrix (respectively, haplotype matrix) is
$1$ if $\sigma_j$ belongs to the $i$-th genotype (respectively,
$i$-th haplotype) and is equal to $0$ otherwise.
In the following we identify rows of a genotype (or haplotype)
matrix with the corresponding genotypes (or haplotypes).
Given a matrix $M$, we denote by $M[\cdot,A]$ (by $M[B, \cdot]$,
respectively) the submatrix of $M$ induced by the set $A$ of columns (by
the set $B$ of rows, respectively).

% Since the columns of the matrices are indexed by the characters and the rows
% are indexed by the genotypes or the haplotypes, sometimes we will use
% interchangeably a character and the column associated with it, or a
% genotype/haplotype and the row indexed by it.

Given a genotype or haplotype matrix $M$ over $\Sigma$, we will
say that a subset $\Sigma_1$ of $\Sigma$ is a \emph{linearly dependent}
set of characters (or, simply, a \emph{dependent} set of characters)
in matrix $M$ if there exists a non-empty subset
$\Sigma_2$ of $\Sigma_1$ such that, for each row $i$,
$\oplus_{\sigma\in\Sigma_2} M[i,\sigma]= 0$.
Otherwise it is called \emph{linearly independent} (or simply
\emph{independent}).

While solving the PPXH problem, we can restrict our attention to a
maximal independent subset of characters as
stated in the following lemma.
\begin{Lemma}
\label{lemma:problema-ridotto-colonne-indipendenti} Let $X$ be a
xor-genotype matrix and $H$ be a haplotype matrix over the same
character set $\Sigma$. Let $\Sigma_1$ be a maximal independent
subset of $\Sigma$ in  $X$. Then, $H$ resolves $X$ if and only if
$H[\cdot,\Sigma_1]$ resolves $X[\cdot, \Sigma_1]$
\end{Lemma}
\begin{proof}
The \emph{only-if} part is obviously true because $H[\cdot, \Sigma_1]$ and
$X[\cdot, \Sigma_1]$ are two submatrices of $H$ and $X$ respectively.
The \emph{if} part can be proved by constructing a feasible
solution $H$ for $X$ from the smaller solution $H[\cdot, \Sigma_1]$ for
$X[\cdot, \Sigma_1]$ (for simplicity we will refer to the two submatrices
respectively as $H'$ and $X'$).
For each character $\alpha \in \Sigma \setminus \Sigma_1$, since
$\Sigma_1\cup\{\alpha\}$ is dependent there exists a non-empty subset
$\Sigma_\alpha$ of $\Sigma_1$ such that, for each  genotype $x$, $X[x, \alpha]=
\oplus_{\sigma\in\Sigma_\alpha}  X'[x, \sigma]$.
Set the entry $H[x, \alpha]$ to
$\oplus_{\sigma\in\Sigma_\alpha}  H'[x, \sigma]$.

We claim that $H$ resolves $X$. Since  $H'$ resolves
$X'$, it suffices to prove that for each  character $\alpha \in
\Sigma \setminus \Sigma_1$, $H[h_1, \alpha] \oplus H[h_2, \alpha] =
X[x, \alpha]$, for some pair of haplotypes $h_1, h_2$.
We already know that  for each genotype $x'$ of $X'$, there is a pair
$(h'_1, h'_2)$ of haplotypes in $H'$ that  resolves $x'$.
Notice that $X[x, \alpha]=\oplus_{\sigma\in\Sigma_\alpha}
 X'[x, \sigma]$ since $\Sigma_1$ is a maximal subset of
 independent characters of $X$.
Since $H'$ resolves $X'$, $\oplus_{\sigma\in\Sigma_\alpha}
 X'[x, \sigma] = \oplus_{\sigma\in\Sigma_\alpha} (H'[h_1, \sigma]
\oplus H'[h_2, \sigma])$.
Moreover, by the
associativity of  $\oplus$,
$\oplus_{\sigma\in\Sigma_\alpha} (H'[h_1, \sigma]
\oplus H'[h_2, \sigma]) =
(\oplus_{\sigma\in\Sigma_\alpha}  H'[h_1, \sigma])
\oplus (\oplus_{\sigma\in\Sigma_\alpha}   H'[h_2,\sigma])$.
Finally, by our construction of the columns of $H$ corresponding to characters in
$\Sigma \setminus \Sigma_1$,
$(\oplus_{\sigma\in\Sigma_\alpha}  H'[h_1, \sigma])
\oplus (\oplus_{\sigma\in\Sigma_\alpha}   H'[h_2,\sigma])
=H[h_1, \alpha] \oplus H[h_2, \alpha]$, hence completing the proof.
\end{proof}

Notice that, given a $n \times m$ xor-genotype matrix $X$,  a
maximal subset of independent characters in $X$ can be extracted
by applying the Gauss-elimination algorithm on the matrix $X$
in $O(nm^2)$ time.
Observe that the proof of
Lemma~\ref{lemma:problema-ridotto-colonne-indipendenti} is constructive and shows how
to compute efficiently a solution  $H$ for $X$   from a solution
$H[\cdot, \Sigma_1]$ for $X[\cdot, \Sigma_1]$.
 We can
introduce another simplification of the instance which can be
performed efficiently.  It affects the construction of the
xor-graph and  allows an efficient reconstruction of an optimal
xor-graph for the general instance, given a xor-graph for the
reduced or simplified instance.

\begin{Lemma}
\label{lemma:problema-ridotto-carattere-1-genotipo}
Let $X$ be an instance of PPXH, and let $\alpha$ be a character of $X$
such that there exists exactly one genotype $x\in X$ with
$\alpha\in x$.
Then there exists an optimal xor-graph $\solG$   for $X$ such that
there is a vertex $v$ of $\solG$ with exactly one edge $e$
incident on $v$ and $\lambda(e)=x$.
\end{Lemma}
\begin{proof}
Let $\solG$ be an optimal xor-graph $\solG$   for $X$. Since $\alpha$
appears in only one genotype in $X$, there is exactly one edge $e$
of $\solG$ such that  $\alpha\in\lambda(e)$.
By Lemma~\ref{lem:cut-symbol}
removing $e$ from $\solG$ results in a
bipartition $\{H_\alpha,\bar{H_\alpha}\}$ where $H_\alpha$ consists of the
haplotypes containing $\alpha$.
Let $v\in H_\alpha$ and $w\in
\bar{H_\alpha}$ be the two endpoints of  $e$,
and let $D$ be the set of vertices of
$H_\alpha$ adjacent to $v$.
Change each haplotype in $h\in H_\alpha \setminus \{v\}$ to $h\oplus
v\oplus w$, obtaining a new  xor-graph $\solG_1$.

By construction,  $\solG_1$ has set of edges
$E_1 = E \setminus \{(v,d) \mid d \in D\} \cup \{(w, d) \mid d \in D\}$.
Indeed, let $e=(v,d)$ be any edge of $\solG$ connecting $v$ with a vertex
$d\in D$; in  $\solG_1$ there is an edge $f=(w,d)$ such that
$\lambda(e)=\lambda(f)$.
It is immediate to notice that $\solG$ and $\solG_1$ have the same number of
vertices, therefore $\solG_1$ is optimal and satisfies the statement of the lemma.
\end{proof}

Also the proof of Lemma~\ref{lemma:problema-ridotto-carattere-1-genotipo} is
constructive and can be exploited directly in an algorithm to simplify the
instance of the problem. More precisely,
the removal of a genotype and a character as stated by
Lemma~\ref{lemma:problema-ridotto-carattere-1-genotipo} can be
repeated until all characters appear in at least two genotypes (or
we obtain the special case of an instance containing only one
genotype; in such case the optimal solution is trivially made by
two haplotypes.) Following the same idea, if the set of characters
is linearly dependent, we can extract a maximal subset of linearly
independent characters. Moreover the executions  of the two
reductions can be intertwined until none of those reductions can
be performed.
% Notice that all reduced matrices must have at least
% as many rows as columns.

An instance $X$ of PPXH is called  \emph{reduced} if (i) $X$
consists of only one genotype, or the two following conditions are
satisfied: (ii a) the set of characters of $X$ are
independent and (ii b) each character appears in at least two
genotypes. Lemmas~\ref{lemma:problema-ridotto-colonne-indipendenti}
and~\ref{lemma:problema-ridotto-carattere-1-genotipo}   justify the
fact that we will assume in the rest of the paper that all
instances are reduced, as the reduction process can be performed
efficiently, and we can easily compute a solution of the original
instance given a solution of a reduced instance (see
Algorithm~\ref{alg:reduction} for a more detailed description of the
reduction process).

\begin{algorithm}[ht!]
\caption{The reduction step}
\label{alg:reduction}
\KwData{a xor-genotype matrix $X$}
\KwResult{a reduced xor-genotype matrix associated with the input matrix
  $X$}
\Repeat{$D$ and $A$ are both empty}{%
  $C\gets$ subset of linearly independent columns of $X$ obtained by the
  Gauss-elimination algorithm\;
  $D\gets X\setminus C$\;
  $A\gets $ set of symbols appearing in exactly one genotype in $X$\;
  Remove from $X$ all columns in $D$ and all rows in $A$\;
}
\Return{$X$}\;
\end{algorithm}

The reduction process leads us to an important lower bound on the size
of the optimum.

\begin{Lemma}\label{lem:lb-m}
Let $X$ be a reduced genotype matrix having $n$ rows and $m$ columns.
Then any haplotype matrix $H$ resolving $X$ has at least $m+1$ rows.
\end{Lemma}
\begin{proof}
Let $\solG$ be a xor-graph for $X$.
By Lemma~\ref{lem:cut-symbol}, each character $\alpha$ induces a cut in
graph $\solG$.
Each cut can be represented as $n$-bit binary vector $c_\alpha$ in
which each element $c_\alpha[i]$ is equal to 1 if and only if the
genotype $x_i$ belongs to the cut.
Clearly, such vector is precisely the column vector corresponding to
character $\alpha$ of matrix $X$.
Thus, since the characters are independent, also the
family of the cuts (represented as binary vectors) induced by the set of
characters is linearly independent.
By Theorem~1.9.6 of~\cite{DIE05B} all
connected graphs with  $m$ independent cuts have at least
$m+1$ vertices.
\end{proof}

%\subsection{Reduction of the Instance Size}
%\label{sec:kern}

As a consequence of Lemma~\ref{lem:lb-m},  in a \emph{reduced} xor-genotype
matrix, the number of rows is greater than or
equal to the number of columns. In fact, in any matrix, the number
of linearly independent columns is equal to the number of
linearly independent rows and, clearly, is bounded by the minimum
between the number of columns and the number of
rows.

The process of reducing a xor-genotype matrix by restricting
ourselves to a maximal subset of independent characters is an
application of the \emph{kernelization} technique for designing a
fixed-parameter algorithm~\cite{ParameterizedComplexity}. The
technique consists of reducing the original instance to a new
instance whose size depends only on the parameter (in our case the
size of the optimal solution.) The size of the reduced
xor-genotype matrix is clearly bounded by a polynomial function of
the optimum $k$ since at most $O(k^2)$ distinct genotypes can be
generated by $k$ distinct haplotypes and, by the previous
consideration, the number of columns is less than the number of
rows. As a result, the number of entries of a reduced xor-genotype matrix is
bounded by $O(k^4)$.
% \notaestesa{YP}{(Anche se non serve) La dimensione e' limitata a
% $O(k^3)$ per il lemma \ref{lem:lb-m}}

\section{Algorithms for Restricted Instances}
\label{sec:poly}

In this section we investigate two restrictions of the PPXH
problem obtained by  bounding    the number of characters that can
appear in each genotype and   the number of  genotypes where a
character can occur. Those restrictions are summarized by the
following formulation.

\begin{Problem} \textsc{Constrained Pure Parsimony Xor Haplotyping (PPXH($p,q$)).}
The instance consists of  a set $X$ of xor-genotypes, where each
xor-genotype $x\in X$ contains   at most $p$ characters, and each
character appears in at most $q$ xor-genotypes. The goal is to
compute a minimum cardinality set $H$ of haplotypes that resolves
$X$.
We use the symbol $\infty$ when one of parameters $p$ or $q$ is unbounded.
\end{Problem}

More precisely we will present efficient algorithms for  the case
when each character is contained in at most two xor-genotypes
(PPXH($\infty,2$)) and the case that each genotype consists of at
most two characters (PPXH($2,\infty$)).

\subsection{A Polynomial Time Algorithm for PPXH($\infty,2$)}

The structure of the cycles in a xor-graph characterizes the solutions for the
PPXH($\infty,2$) problem as stated in the following Lemma.

\begin{Lemma}\label{lem:case-2-no-cont-cycle}
Let $X$ be a reduced instance of PPXH($\infty, 2$), let $\solG$ be an
optimal xor-graph for $X$, and let $e$ be an edge of $\solG$.
Then $e$ belongs to exactly one simple cycle of $\solG$.
\end{Lemma}

\begin{proof}
Assume to the contrary that an edge $e$ belongs to two cycles
$C_1$ and $C_2$. Notice that the three sets $C_1 \setminus C_2$,
$C_2 \setminus C_1$, $C_1\cap C_2$ are pairwise
disjoint and not empty. Let $d$ be any element of $\lambda(C_1 \cap C_2)$. By
Lemma~\ref{lem:0-sum}, $\oplus(\lambda(C_1 \setminus C_2)) =
\oplus(\lambda(C_2 \setminus C_1)) = \oplus(\lambda(C_1 \cap
C_2))$. Consequently there exist three distinct edges $e_1\in C_1 \setminus
C_2$, $e_2 \in C_2 \setminus C_1$, $e_3 \in C_1 \cap C_2$ such
that $\lambda(e_1)$, $\lambda(e_2)$, $\lambda(e_3)$ all contain
$d$, which  contradicts  the fact that there
are only two genotypes containing $d$.  By the first part of the proof,
 we have now to prove that $e$
belongs to at least a cycle of $\solG$.

Assume to the contrary
that $X$ is a smallest counterexample, that is no such xor-graph
exists for $X$, while such graph exists for all reduced
instances with fewer genotypes, and let $\solG$ be any optimal
xor-graph for $X$. Since there is an edge that does not belong to any
cycle of $\solG$, there is  a character $a$ such that both edges
$e_1$ and $e_2$ containing $a$ do not belong to any cycle. Notice that two
such edges must exists, since the instance is reduced.
Let us denote by $\alpha$, $\beta$ and $\gamma$ respectively the sets
$\lambda(e_1)\cap \lambda(e_2)$, $\lambda(e_1)\setminus \lambda(e_2)$, $\lambda(e_2)\setminus \lambda(e_1)$.

Compute a new reduced instance $X_1$ from $X$ by removing
the xor-genotypes $e_1$  and $e_2$, and adding a
new genotype $x_c=\beta \cup \gamma =  e_1 \oplus e_2$.
Clearly $X_1$ is a reduced instance of PPXH($\infty, 2$) smaller than $X$, therefore $X_1$ admits
an optimal graph $\solG_1$ where all edges are in some cycle. Let
us consider the unique cycle $C$ of $\solG_1$ containing the edge $e=(u,v)$, with
$\lambda(e)=x_c$.
Now, starting from $\solG_1$, compute a
xor-graph $\solG_2$ for instance $X$, by adding to $\solG_1$
a new vertex $w$, two edges $e'_1=(u,w)$,
$e'_2=(w,v)$, so that $\lambda(e'_1)=e_1$ and $\lambda(e'_2)=e_2$,
and by removing edge $e$.
The graph $\solG_2$ is a xor-graph of $X$ as $x_c = e_1 \oplus e_2$.
Clearly the newly obtained
graph $\solG_2$ is a xor-graph for $X$ satisfying the
requirements of the lemma and  $\solG_2$ contains one more
vertex than $\solG_1$.

We have to prove that $\solG_2$ is optimal, therefore assume that $\solG_2$ is not
optimal and let $\solG_*$ be an optimal xor-graph for $X$, that is $\solG_*$
has no more vertices than $\solG_1$. It is immediate to notice that
contracting each of $e_1$ and $e_2$ into single vertices result in a xor-graph
that is a solution of $X_1$ with fewer vertices than $\solG_1$, hence violating
the optimality of  $\solG_1$.
\end{proof}

Since the optimal xor-graph is connected
(Lemma~\ref{lem:opt-connected}) and consists of a set of edge-disjoint cycles
(Lemma~\ref{lem:case-2-no-cont-cycle}), the size of the optimum solution is
equal to $|X| + 1 - |\cal{C}|$, for $|X|$ the number of genotypes or
edges of the graph and $|\cal{C}|$ the number of simple cycles of the
graph, since any set of $|\cal{C}|$  simple cycles on a graph with at most
$|X| - |\cal{C}|$ must share at least an edge.

Algorithm~\ref{alg:.,2} solves the PPXH($\infty,2$) problem by computing the
set $\mathcal{C}$ of all simple cycles of an optimal xor-graph. In fact
Lemma~\ref{lem:case-2-no-cont-cycle} allows us to introduce a binary
relation $R$ between genotypes, where two genotypes are related if and only if
they share a common character. By Lemma~\ref{lem:case-2-no-cont-cycle} any two
genotypes (or edges of the xor-graph) that are related must also belong to the
same simple cycle. It is immediate to notice that the partition of the edges
of the xor-graph into simple cycles is equal to the most refined partition of
edges such that any two edges sharing a common character belong to the same
set of such partition. In fact Algorithm~\ref{alg:.,2} computes exactly the
closure of $R$.

% An algorithm solving PPXH($\infty,2$) consists of building the optimal
% xor-graph which, in this case, means computing a most refined
% partition of the genotypes $X$ into nonempty sets $X_1,\ldots X_k$
% where $\oplus X_i=\emptyset$ for each $X_i$.
% Each element $X_i$ of the partition, since it is a subset of genotypes
% whose sum is equal to $\emptyset$, will correspond to a cycle of the
% optimal xor-graph.
% In the following let
% us show an algorithm that computes such a partition in polynomial
% time. The algorithm builds iteratively an auxiliary graph whose
% vertices are the input genotypes $X$ and where two genotypes are
% adjacent in the graph if and only if   they share a common
% character. Since $\oplus X_i=\emptyset$ for a set of the partition
% of $X$,  for each character $c$ either none or both of the
% genotypes containing $c$ must belong to $X_i$, which implies that,
% given two genotypes that are adjacent in the
% auxiliary graph, either none or both of them must be in $X_i$.
% This fact immediately implies that we can easily obtain an optimal
% xor-graph composed by the edge-disjoint cycles corresponding to the
% connected components of the auxiliary graph.
\begin{algorithm}[ht!]
\label{alg:.,2}
\KwData{a set $X$ of reduced xor-genotypes}
\KwResult{an optimal solution $H$ for $X$}
$\mathcal{C}\gets \emptyset$\;
\While {$X \neq \emptyset$}{%
  $x \gets$ any element of $X$\;
  $C\gets \{x\}$\;
  \Repeat {$D=\emptyset$}{%
    $D\gets $ the set of genotypes in $X\setminus C$ sharing at least one element with
    some genotypes in $C$\;
    $C\gets C\cup D$\;
  }
  Add $C$ to $\mathcal{C}$\;
  Remove all genotypes in $C$ from $X$\;
}
\ForEach {$C\in\mathcal{C}$}{%
  Transform $C$ into a cycle\;
}
Build a graph $H$ from $\mathcal{C}$, so that all cycles in $\mathcal{C}$ share a
common vertex\;
\Return{$H$}
\caption{The algorithm for PPXH($\infty,2$)}
\end{algorithm}

\subsection{A Polynomial Time Algorithm for PPXH($2,\infty$)}

For simplicity's sake we will assume that the instance of the problem is
a genotype matrix $X$ and the desired output is a haplotype matrix $H$.

We remember that in both matrices the columns are indexed by
characters therefore we will denote by $X[\cdot,\sigma]$ (respectively
$H[\cdot,\sigma]$) the column of $X$ (resp. $H$) indexed by the character $\sigma$.
The algorithm is based on
Lemmas~\ref{lemma:problema-ridotto-colonne-indipendenti} and
\ref{lem:lb-m}.

In fact we will first compute a largest set $\Sigma_1$ of
independent characters in $X$. Moreover for each character $\alpha
\in \Sigma \setminus \Sigma_1$ we determine the subset $ \Sigma_{\alpha}$ of
$\Sigma_1$ such that $X[\cdot,\alpha]=\oplus_{\sigma \in \Sigma_{\alpha}}
X[\cdot,\sigma]$. Notice that this step can be carried over by a simple
application of the Gauss-elimination algorithm.

Let $X'$ be the submatrix $X[\cdot,\Sigma_1]$.
An optimal solution of the instance $X'$
is the matrix $H'$ containing $|\Sigma_1|+1$ rows. More
precisely the $i$-th row of $H'$, for $1\le i\le
|\Sigma_1|$, consists of all zeroes, except for the $i$-th column
(where it contains $1$). The last row contains only zeroes.

Clearly $H'$ resolves $X'$.
In fact it is immediate to notice that each row of $X'$
contains at most two $1$s, as the same property holds for $X$,
therefore for each  row $r$ of $X'$ there are two rows of
$H'$ resolving $r$. The optimality of such solution is a
direct consequence of Lemma~\ref{lem:lb-m}.
Clearly
$H'$ is not a feasible solution of the original instance
$X$, but such a feasible solution $H$ can be easily computed from
$H'$ by adding, for each character $\alpha\in
\Sigma \setminus \Sigma_1$, a column equal to $\oplus_{\sigma \in
\Sigma_{\alpha}} H'[\cdot,\sigma]$ (where $\Sigma_\alpha$ satisfies $X[x, \alpha]=
\oplus_{\sigma\in\Sigma_\alpha}  X'[x, \sigma]$ for each  genotype $x$.)
  The matrix $H$ is a feasible
solution as shown in the proof of
Lemma~\ref{lemma:problema-ridotto-colonne-indipendenti}.

\section{Fixed-Parameter Tractability of PPXH}%
\label{sec:fpt}

As observed at the end of Section~\ref{sec:xorgraph}, the reduction
of an instance of the unrestricted PPXH problem lead us to a
fixed-parameter algorithm (where the parameter is the optimum).
Moreover we can observe that there exists another fixed-parameter
algorithm for the unrestricted PPXH problem without using the reduction
of input instance.
Let $H$ be a set of haplotypes and let $X$ be a set of
genotypes resolved by $H$; in the following we will denote  $\card{X}$ by $n$.
Since $H$ can resolve at most $\binom{\card{H}}{2}$
genotypes, $n\ge \card{H} \ge \sqrt{2n}$. In other words, if $k$ is the size of
the minimum-cardinality set of haplotypes resolving $X$, $n\in O(k^2)$.

The number of the possible graphs with at most $n+1$ vertices and
exactly $n$ edges is no more than $ 2 ^ {2n\log_2{(n+1)}} =
(n+1)^{2n}$ which, by our previous observation, is $O ( k^{4k^2} )$,
i.e.~a function dependent only on $k$. The time needed to
check if one of such graphs is a xor-graph for $H$ is clearly
polynomial in $n$ and thus we can immediately derive a fixed-parameter
algorithm to find an optimal xor-graph for $X$.

The time complexity of the algorithm is well beyond what is deemed
acceptable in practice, therefore we propose a more efficient
algorithm that is based on the matrix representation of genotypes
and haplotypes.

In the following we will assume that the genotype matrix $X$ is
reduced, and that $X$ has $n$ rows and $m$ independent columns, and
that we are looking for a haplotype matrix $H$ with at most $k$
distinct rows that resolves $X$.
The basic idea of our algorithm is to enumerate all possible haplotype
matrices.

In the na\"ive approach, testing
if a haplotype matrix resolves a given genotype matrix requires
$O(k^2nm)$ time because each pair of haplotypes has to be
considered and then each resulting genotype has to be searched in
the genotype matrix. Our strategy, instead, is to enumerate all
the haplotype matrices by changing only one haplotype each time,
in such a way that only $k-1$ new pairs of haplotypes must be
considered
 when testing if $H$ resolves the set $X$.

We use Gray codes~\cite{Savage97}   to visit all the haplotype
matrices in such a way that each pair of consecutive matrices
differs by a single bit and, thus, by a single haplotype. More
precisely, we enumerate all $k \times m$ matrices by generating all
$km$-long bit vectors. Indeed,  the bits from position $(i-1)
m+1$ to position $i m$ in a $ km$-long vector give the $i$-th row
of the matrix (for $1 \le i \le k$). The fastest known algorithm
for computing the next vector of a Gray code requires constant
time for each invocation~\cite{BitEhRei76GrayCode}.

Observe that the na\"ive algorithm  requires
$O(nm)$ time to test if there is a genotype in matrix $X$
resolved by a pair of haplotypes. By   representing the set of the row
vectors of matrix $X$ as a
\emph{binary trie}~\cite{Fred60Trie}, the
time required to get the index of the   row  containing a
$m$-long binary vector is reduced to $O(m)$.

The details of the fixed-parameter algorithm are given in the
Algorithm~\ref{alg:fpt}, where we also use some additional data structures:
the array \emph{ResolvedByHowMany} which associates with each genotype the
number of pairs of haplotypes resolving such genotype, and
\emph{ListResolvedGenotypes} which associates with each haplotype $h$ a list  of the
relevant pairs of haplotypes in which $h$ is involved. In fact, the
elements of the lists in \emph{ListResolvedGenotypes} are triples $(h_1, h_2, x)$
where $(h_1, h_2)$ is a pair of haplotypes resolving $x$.

\begin{algorithm}[htb!]
\newcommand{\var}[1]{\ensuremath{\text{\emph{#1}}}}
\caption{A fixed-parameter algorithm for PPXH}
\label{alg:fpt}
\KwData{A genotype matrix $X$ defined over a set of $m$ independent characters, and an integer $k$.}
\KwResult{a set $H$ of at most $k$ haplotypes resolving $X$ if it exists, \emph{No} otherwise.}

\lIf{$\binom{k}{2} < n$ or $k<m$}{\label{alg:fpt:bounds-begin}
\Return{\itshape No}\;
}
\lIf{$k>n$}{
  \Return{$H\cup \{h_0\}$}\;
}\label{alg:fpt:bounds-end}
%\Else{
  Build a trie $T$ that stores the xor-genotypes contained in $X$\;
  Let $\var{ListResolvedGenotypes}$ be an array of $k$ initially empty lists\;
  $\var{ResolvedByHowMany} \gets (0,0,\ldots,0)$\;% \tcp*{array of $n$ zeroes}
  $\var{TotalResolvedG} \gets 0$\;
  \ForEach{binary matrix $H$ in Gray code}{\label{alg:fpt:main-for-begin}
    \lIf{$H$ is the matrix containing only zeros}{continue to the next matrix\;}
    $\var{ChangedRow}\gets$ index of the row changed from the previous iteration\;
    \tcc{\small Update state of xor-genotypes resolved by changed
      haplotype}
    \ForEach{entry $(h_1, h_2, x)$ of
      $\var{ListResolvedGenotypes}[\var{ChangedRow}]$}{\label{alg:fpt:first-block-begin}
      Remove $(h_1, h_2, x)$ from $\var{ListResolvedGenotypes}[h_1]$ and
      $\var{ListResolvedGenotypes}[h_2]$\;
      $\var{ResolvedByHowMany}[x] \gets \var{ResolvedByHowMany}[x]-1$\;
      \lIf{$\var{ResolvedByHowMany}[x]=0$}{$\var{TotalResolvedG} \gets \var{TotalResolvedG} - 1$\;
      }
    }\label{alg:fpt:first-block-end}
    \tcc{\small Look for genotypes resolved by the new haplotype}
    \For{$r \gets 1$ \KwTo $k$}{\label{alg:fpt:second-block-begin}
      \If{$l$ is the index returned by the lookup of the vector $H[r, \cdot] \oplus H[\var{ChangedRow}, \cdot]$ in $T$}{
        \lIf{$\var{ResolvedByHowMany}[l]=0$}{%
          $\var{TotalResolvedG} \gets \var{TotalResolvedG} + 1$\;
        }
        $\var{ResolvedByHowMany}[l] \gets \var{ResolvedByHowMany}[l] +1$\;
        Add $(r, \var{ChangedRow}, l)$ to $\var{ListResolvedGenotypes}[r]$ and to
        $\var{ListResolvedGenotypes}[\var{ChangedRow}]$\;
      }
    }\label{alg:fpt:second-block-end}
    \If{$\var{TotalResolvedG} = n$}{
      \tcc{\small all genotypes are resolved}
      Remove from $H$ all duplicate rows\;
      \Return{$H$}\;
    }
  }\label{alg:fpt:main-for-end}
  \Return{\itshape No}\;
%}
\end{algorithm}

Notice that the outermost foreach loop (lines 6--25) iterates
$2^{km}$ times, while the for loop at lines
\ref{alg:fpt:second-block-begin}--\ref{alg:fpt:second-block-end}
iterates $k$
times. Each iteration of the latter loop consists of a lookup in a
trie (which can be done in $O(m)$ time) and updating in constant time
some arrays and lists.
Since each list can contain at most $k$ elements, the
time required for each iteration of the outermost loop is $O(km)$,
resulting in an overall $O(nm+2^{k^2}km)$ time complexity.

\section{An Approximation Algorithm}
\label{sec:approx}

We present a simple approximation algorithm, detailed as
Algorithm~\ref{alg:apx}, which guarantees for a reduced instance $X$ of
PPXH an approximation factor $l$, where $k$ is the maximum number of
xor-genotypes where each character appears.

Initially the set $H$ of haplotypes computed by the algorithm contains only
the null haplotype. While the set of genotypes is not empty, pick a character
$\alpha$ that appears in at least a genotype, move to $H$ all genotypes
containing $\alpha$, and remove from $X$ all genotypes that are solved
by a pair of haplotypes in $H$.
Clearly the final set of haplotypes $H$ solves the set of genotypes $X$.

\begin{algorithm}[ht!]
\label{alg:apx}
\KwData{a set $X$ of xor-genotypes over alphabet $\Sigma$}
$H \gets \{h_0\}$\;
\While {$\Sigma \neq \emptyset$}{%
  $\alpha \gets$ any character in $\Sigma$\;
  \ForEach{$x \in X\text{ s.t. } \alpha \in x$}{%
    add to $H$ the genotype $x$\;
  }
  Remove from $X$ all genotypes that contains the character $\alpha$\;
  Remove $\alpha$ from $\Sigma$\;
  }

\KwResult{$H$}
\caption{The approximation algorithm}
\end{algorithm}

The proposed algorithm returns a solution of size at most $l$ times
larger than the optimum which, by Lemma~\ref{lem:lb-m}, is at least
$\card{\Sigma}+1$.
Our algorithm starts with a partial solution $H$ containing only the
null haplotype, and at each iteration adds at most
$l$ haplotypes to the solution $H$, as $l$ is the maximum number of
genotypes containing any character. Since there can be at most $\card{\Sigma}$
steps, $\card{H}\le l\card{\Sigma}+1$.

Clearly the approximation ratio is at most $(l|\Sigma|+1)/(|\Sigma|+1)\le l$,
completing the proof.

\section{Solving PPXH by a Heuristic Method}
\label{sec:heur}
% con sperimentazione
In this section we propose a heuristic algorithm to build a near
optimal xor-graph for an input matrix $X$ of genotypes. Observe
that an optimal  xor-graph for $X$ is a  graph having the minimum-cardinality
vertex set and where each edge is uniquely labeled by a genotype $X$.
By Lemma~\ref{lem:0-sum},
a cycle  of the xor-graph consists of a subset  $X'$ of the input genotypes such
that $\oplus X'=\emptyset$. Consequently we will call a  subset  $X'$
with $\oplus X'=\emptyset$ a \emph{candidate cycle}.

The basic idea that guides our heuristic is first to select a
subset of the candidate cycles of  $X$  and then to build a
labeled graph (a xor-graph)  where the selected  candidate cycles
are actual cycles. The procedure successively iterates over
the genotypes that are not yet successfully realized in the
xor-graph.

A related problem is the one called \emph{Graph Realization
(GR)}~\cite{Tutte60}, which consists of  building a graph
given its fundamental cycles. We recall that the set $\mathcal{C}$ of
fundamental cycles of a graph $G$
with respect to a fixed spanning tree $T$ of $G$, is defined as
$\mathcal{C} = \{ \text{\emph{the unique cycle of }}T\cup \{e\} \mid e \in
E(G)\setminus E(T)\}$ (see e.g.~\cite{DIE05B}, pag.~26).
More precisely,
the Graph Realization problem can be formally stated as follows~\cite{Tutte60}.
Given two disjoint sets $T$ and
$C$, the input of the GR problem is a family $F$ of subsets of
$T\cup C$ such that (i) for each set $F_i$ of the family $F$, $F_i
\cap C= \{c_i\}$, and (ii) for each pair of subset $F_i$ and $F_j$
of $F$, $F_i \cap F_j \cap C = \emptyset$. The GR problem consists
of finding a labeled graph $G=(V,E)$ (if such a
graph exists) which \emph{realizes} $F$, that is there is a
bijection between the set $T$ and a spanning tree of $G$, and the
elements of each set $F_i$ label exactly the edge set of a
(simple) cycle of $G$.

In the case that we have selected a set of candidate cycles
which are fundamental cycles of a graph $G$, an immediate application of any
algorithm solving the GR problem (two almost linear time algorithms
exist~\cite{Bixby88,Fujishige80}), gives a xor-graph resolving all those
candidate cycles.
We have been inspired by those algorithm for GR to develop our heuristic. We
denote by $G(F)$ a graph realization $G$ of a family of sets $F$.

The heuristic  procedure transforms a $r \times c$ genotype matrix $X$ into an
instance of GR    as described in the following two main steps. In
the first step,  the set $T$ is defined as a maximal   subset
$T=\{x_{j_1},\ldots , x_{j_c}\}$ of linearly independent input
genotypes of $X$. This means that any  other input genotype $x_i$
can be expressed as a linear combination $ \alpha_{i,1} x_{j_1}
\oplus \ldots \oplus \alpha_{i,c} x_{j_c}$ of the genotypes in $T$.
Then  $C = \{c_1, \ldots c_{r-c}\}$ is defined as consisting of
the set of genotypes not in $T$. In a second step,  the family of
subsets of $T \cup C$ giving an instance of the GR  is built  by
building  sets $F_i$ such that $F_i = \{c_i\}\cup \{x_{j_l}\in T
\mid \alpha_{i,l}=1\}$. Informally, $F_i$ consists of $c_i$ and the
unique set $P_i \subseteq T$, such that $\oplus P_i= \{c_i\}$. An
immediate consequence of our definition is that $\oplus
F_i=\emptyset$, therefore $F_i$ is, by  definition, a candidate cycle.

Computing the set $T$ from $X$ is simply a matter of running the
Gauss elimination algorithm on $X^T$ (that is the transpose matrix of $X$.)
The family $F$ can be easily
inferred by computing the coefficients $\alpha_{i,1},\ldots,\alpha_{i,c}$ for all $c_i \in C$,
%by solving the system of linear equations $X[T,\cdot]^T \alpha_i = c_i$.
 where the
unknowns $\alpha^i_1 \ldots \alpha^i_r$  are the coefficients of the
linear combination and the binary matrix $M$ is a matrix whose columns
are the xor-genotypes in $I$.

Clearly, the Gauss-elimination procedure applied on the matrix
$X^T$ results in a matrix $R$ whose first $r$ columns form the
identity matrix while the other columns are the vectors of the linear
combination coefficients.

We have a final hurdle, that is to handle the case where the GR does not
exist for the family $F$.
Once the family $F$ is identified, the heuristics computes a
maximal subfamily $F'$ of $F$, so that there exists a GR from
$F'$.
Now, let us detail the construction of the family $F$ giving an
instance of GR. The heuristic starts defining $F'$ as an empty
family and iteratively adding to $F'$ a candidate cycle $F_i$ if
and only if the resulting family admits a Graph Realization.
Clearly, this approach ends with a maximal subset of candidate
cycles that admits a Graph Realization. The two
steps of the  heuristic procedure are then recursively iterated on
the set of xor-genotypes of $X$ that do not label an edge of the
computed Graph Realization.
The details of the procedure are
presented in Algorithm~\ref{alg:heu}.

\begin{algorithm}
\caption{The heuristic  {\itshape Heu}$(X)$}\label{alg:heu}
\KwData{a xor-genotype reduced  matrix $X$
  % whose columns are labeled over alphabet $\Sigma$
} \KwResult{a haplotype matrix $H$ that resolves $X$}
%$H \leftarrow \{ h_0 \}$ \tcp*[l]{$H$ is only composed by the null haplotype}
  $r\gets $ the number of rows of $X$\;
  $c\gets $ the number of columns of $X$\;
  \If{$r=c$}{%(i.e.~if all the rows and the columns are
         %linearly independent),
    \Return{the set consisting of $h_0$ and the canonical haplotypes of $X$}\; \label{alg:heu:lin-ind}
  }
  $R\gets $ output of  Gauss elimination on $X^T$\; \label{alg:heu:gj}
  $T=\{x_1, \ldots, x_c\}\gets$ the independent xor-genotypes labeling the first
  $c$ columns of $R$, and $C=\{x_{c+1}, \ldots x_r\}\gets$ the set of the
  remaining genotypes\;\label{alg:heu:gauss}
  $F\gets \emptyset$\;
  \For{$x_i \in C$}{%
    $Z(x_i) \gets \{ x_i \} \cup \{ x_j \in T \mid$ the element in row
    $j$  and column $i$ of $R$ is equal to $1\}$\;
    \If{$F \cup \{ Z(x_i)\}$ admits Graph Realization}{%
      $F \gets F \cup \{ Z(x_i) \}$\;
    }
  }
  Let $G(F)$ be the Graph Realization of $F$\;
  $H\gets $  the set of vertices of $G(F)$\; \label{alg:heu:adding-corresponding-labelling}
  $v\gets$ a random element of $H$\;
  Each $h\in H$ becomes $h\oplus v$ \tcp*[l]{\small Now $v$ is the null haplotype}
  Remove from  $X$ the genotypes that label any edge of
  $G(F)$\; \label{alg:heu:remove}
  \tcc{\small The instance $X$ is
    reduced before the subroutine {\itshape Heu} is recursively called, and
    the general solution is then obtained as described in the proof of
    Lemma~\ref{lemma:problema-ridotto-colonne-indipendenti}}
  \Return{$H \cup \text{\itshape Heu}(X)$}\;
\end{algorithm}

Let $n$ and $m$ be, respectively, the number of xor-genotypes and
sites. The time complexity of the heuristic is determined by the
time complexity of the Gauss elimination algorithm,
which requires $O(n^2m)$ time because it is called on matrix
$X^T$, and of the Graph Realization algorithm, whose best time
complexity is $O(\alpha(n,m)nm)$, where $\alpha$ is the inverse
Ackermann function. Notice that the Graph Realization algorithm is
repeated at most $n$ times in order to compute a maximal subfamily
$F'$, hence $O(\alpha(n,m)n^2m)$ times. Finally, there is at least
one xor-genotype of $X$ that labels an edge of the Graph
Realization, hence
the total number of iterations is at most $n$, leading to an
overall time complexity $O(\alpha(n,m)n^3m)$.

\subsection{Experimental Results}

We have implemented our heuristic as a C program using the
software \texttt{GREAL}~\cite{GREAL} as a routine to solve the
Graph Realization problem.
The \texttt{GREAL} program implements the algorithm of Gavril and
Tamari~\cite{GavTam83}  even if its time
complexity is  $O(nm^2)$ (opposed to the $O(\alpha(n,m)nm)$ time complexity of
the best known algorithm), since it is still effective for our
purposes.

The experimental analysis of our heuristic is composed of two parts.
In the first part we have applied the algorithm on synthetic instances
to evaluate the quality of the results in terms of cardinality
of the solutions and running time.
In the second part we have assessed the applicability of the heuristic
to some real-world large instances.

\subsubsection{Synthetic Data}
Each synthetic instance has been created starting from a set of
initial haplotypes and then each xor-genotype has been generated as
combination of two haplotypes randomly selected from the initial set.
Notice that such process does not guarantee that every haplotype is
selected to form a genotype.

We have used two different methods to generate the set of initial
haplotypes: (a) pure random generation, and (b) generation under the
neutral model.
The first strategy, pure random generation, selects uniformly sets of
$h$ distinct haplotypes from the set of all binary haplotypes of
length $m$.
The second strategy, generation under the neutral model, uses
the standard Hudson's simulator \texttt{ms}~\cite{Hudson02} to generate
a sample of $h$ haplotypes assuming the neutral model of genetic
variation.
In this case, the sample of haplotypes can contain repeated elements.
Using two different methods to generate the set of initial haplotypes
allows us to verify if the behavior of the heuristic is influenced by
the choice of the initial haplotypes.

The evaluation criteria, in both cases, were (a) the number of distinct
haplotypes computed by our method, and (b) its running time.
In particular, we have considered as main indicator of the quality of
the solutions the ratio ($r$) between the number of distinct haplotypes
of the computed solution and the number of distinct initial haplotypes
selected to generate a genotype of the instance.
We notice that $r$ is only a proxy for the actual approximation ratio (that is
the ratio  between  the number of distinct computed haplotypes and the size of
a optimal solution) achieved by the algorithm, as the number of the selected
haplotypes represents only an
upper bound of the optimum, thus the ratio $r$ might be strictly less than $1$.

Since the outcome of our heuristic can be influenced by
the order of the input genotypes, for each instance we have run the algorithms
on  ten random
permutations of the genotypes, and we have retained only the smallest set of
computed haplotypes.
The running time refers to the total time required by the heuristic on
the 10 permutations of genotypes and has been measured on a standard PC
with 1GB of memory with CentOS Linux 5.

The pure random generation strategy is characterized by three
parameters, namely the number of input genotypes ($n$), the number of
haplotypes ($h$), and the number of characters ($m$).
We have considered 4 different values of the parameter $n$ ($100$, $200$,
$300$, $400$), and we have computed the values of $h$ and $m$ from $n$: in
fact those values are ${n}/4$, ${n}/{3}$, and ${2n}/{3}$.
The maximum size of the test instances (400 genotypes and 233
characters) has been chosen in such a way that repeated tests on
several instances of the same size would be feasible on a normal
computer.
In fact, as discussed below, on average the heuristic required roughly an hour
on the largest instances, therefore any further increase of the instance size
would have made the experimentation impractical.
% Ho eliminato il caso n=50 per mantenere regolare la tabella.

Table~\ref{tab:ris-heu-rnd} reports the average size of the solutions
computed by our heuristic, its average running time,
and the average ratio $r$ on 10 random instances generated for each
choice of the parameters $n$, $h$, and $m$.

% The results for the first class of instances (those generated using the
% first strategy,
% random haplotypes) are presented in Table~\ref{tab:ris-heu}.
% In particular, it reports the average size of the solutions computed by
% our heuristic and the average running times over $10$ instances for each
% choice of the three parameters, namely the number of input
% genotypes($n$), the number of haplotypes ($h$), and the number of
% characters ($m$).
% We have considered 5 different values of the parameter $n$ (50, 100, 200,
% 300, 400), and we have computed the values of $h$ and $m$ from $n$ using
% the following functions: $5\sqrt{n}$, ${n}/{3}$, and ${2n}/{3}$ (the table
% presents the results in this order).

\begin{table*}[t!]
\caption{%
  Results on instances generated using the pure random strategy.
  For each choice of the first three columns, 10 random
  instances were generated.
  The column \emph{average independent characters} reports the average number
  of independent character in the genotype matrix, while column
  \emph{average initial haplotypes} reports the average number of
  distinct haplotypes selected to generate each instance.
  The last two columns report, respectively, the average size of the
  solution computed by our heuristic and the average ratio $r$.}
\label{tab:ris-heu-rnd}
\centering \scriptsize
\begin{tabular}{c c c r@{}l r@{}l r@{}l r@{}l}% r@{}l}
\toprule
\parbox{1.4cm}{\centering number of\\genotypes\\$n$} &
\parbox{1.4cm}{\centering number of\\generated\\haplotypes\\$h$} &
\parbox{1.4cm}{\centering number of\\characters\\$m$} &
\multicolumn{2}{c}{\parbox{1.4cm}{\centering average independent characters} }&
\multicolumn{2}{c}{\parbox{1.4cm}{\centering average initial haplotypes} }&
\multicolumn{2}{c}{\parbox{1.4cm}{\centering average result size} }&
\multicolumn{2}{c}{\parbox{1.4cm}{\centering average ratio\\$r$} }\\
\midrule
100     & 25    & 25    &\hspace{4.3mm}23&.70   & \hspace{4.3mm}25&.00  & \hspace{4.3mm}25&.90  & \hspace{5mm}1&.04\\
        &       & 33    & 24&.00        & 25&.00        & 25&.00        & 1& \\
        &       & 66    & 24&.00        & 25&.00        & 25&.00        & 1& \\
\cmidrule(lr){2-11}
        & 33    & 25    & 25&.00        & 32&.90        & 51&.60        & 1&.57\\
        &       & 33    & 31&.50        & 32&.80        & 33&.20        & 1&.01\\
        &       & 66    & 32&.00        & 33&.00        & 33&.00        & 1& \\
\cmidrule(lr){2-11}
        & 66    & 25    & 25&.00        & 63&.00        & 87&.30        & 1&.39\\
        &       & 33    & 33&.00        & 63&.30        & 87&.20        & 1&.38\\
        &       & 66    & 62&.70        & 63&.90        & 63&.80        & 1& \\
\midrule
200     & 50    & 50    & 48&.70        & 50&.00        & 50&.90        & 1&.02\\
        &       & 66    & 49&.00        & 50&.00        & 50&.00        & 1& \\
        &       & 133   & 49&.00        & 50&.00        & 50&.00        & 1& \\
\cmidrule(lr){2-11}
        & 66    & 50    & 50&.00        & 65&.80        & 96&.20        & 1&.46\\
        &       & 66    & 64&.50        & 65&.90        & 66&.20        & 1& \\
        &       & 133   & 64&.80        & 65&.80        & 65&.80        & 1& \\
\cmidrule(lr){2-11}
        & 133   & 50    & 50&.00        & 126&.90       & 185&.80       & 1&.46\\
        &       & 66    & 66&.00        & 126&.20       & 186&.10       & 1&.47\\
        &       & 133   & 126&.60       & 128&.10       & 127&.70       & 1& \\
\midrule
300     & 75    & 75    & 73&.70        & 75&.00        & 75&.80        & 1&.01\\
        &       & 100   & 74&.00        & 75&.00        & 75&.00        & 1& \\
        &       & 200   & 73&.90        & 74&.90        & 74&.90        & 1& \\
\cmidrule(lr){2-11}
        & 100   & 75    & 75&.00        & 99&.80        & 149&.80       & 1&.50\\
        &       & 100   & 98&.60        & 99&.90        & 100&.00       & 1& \\
        &       & 200   & 98&.80        & 99&.80        & 99&.80        & 1& \\
\cmidrule(lr){2-11}
        & 200   & 75    & 75&.00        & 190&.80       & 285&.90       & 1&.50\\
        &       & 100   & 100&.00       & 191&.10       & 284&.60       & 1&.49\\
        &       & 200   & 188&.20       & 190&.30       & 189&.20       & 0&.99\\
\midrule
400     & 100   & 100   & 98&.50        & 100&.00       & 100&.90       & 1&.01\\
        &       & 133   & 98&.90        & 99&.90        & 99&.90        & 1& \\
        &       & 266   & 98&.90        & 99&.90        & 99&.90        & 1& \\
\cmidrule(lr){2-11}
        & 133   & 100   & 100&.00       & 132&.80       & 194&.90       & 1&.47\\
        &       & 133   & 131&.40       & 132&.80       & 133&.00       & 1& \\
        &       & 266   & 131&.80       & 132&.80       & 132&.80       & 1& \\
\cmidrule(lr){2-11}
        & 266   & 100   & 100&.00       & 254&.70       & 385&.30       & 1&.51\\
        &       & 133   & 133&.00       & 253&.60       & 384&.40       & 1&.52\\
        &       & 266   & 251&.90       & 253&.50       & 252&.90       & 1& \\
\bottomrule
\end{tabular}
\end{table*}

The second strategy, generation under the neutral model, is
characterized by the three parameters $n$, $m$, and $\rho$, where $n$ is
the number of genotypes, $m$ is the number of characters, and $\rho$ is
the crossover (or recombination) rate of the Hudson's program.
The size of the initial sample of haplotypes has been set equal to the
number $n$ of genotypes.
Since the sample can contain several copies of the same haplotype, the
number of distinct haplotypes randomly selected to form a genotype has
been significantly lower than the number of genotypes for almost all of
the generated instances.

We considered 30 instances for each choice of the parameters $(n, m,
\rho)$ with $n \in \{ 50, 75, 100\}$, $m \in \{ 50, 75, 100\}$, and
$\rho \in \{ 0, 8, 16, 24 \}$.
As for the previous dataset, Table~\ref{tab:ris-heu-hud} reports the
average size of the solution computed by our heuristic, its average running
time, and the average ratio $r$.

\begin{table*}[t!]
\caption{%
  Results on instances generated using the neutral model.
  For each choice of the first three columns, 30 random
  instances were generated.
  The column \emph{average independent characters} reports the average number
  of independent character in the genotype matrix, while column
  \emph{average~initial haplotypes} reports the average number of
  distinct haplotypes selected to generate each instance.
  The last two columns report, respectively, the average size of the
  solution computed by our heuristic and the average ratio $r$.}
\label{tab:ris-heu-hud}
\centering \scriptsize
\begin{tabular}{c c c r@{}l r@{}l r@{}l r@{}l}% r@{}l}
\toprule
\parbox{1.4cm}{\centering number of\\genotypes\\$n$} &
\parbox{1.4cm}{\centering number of\\characters\\$m$} &
\parbox{1.4cm}{\centering recombination\\rate\\$\rho$} &
\multicolumn{2}{c}{\parbox{1.4cm}{\centering average independent characters} }&
\multicolumn{2}{c}{\parbox{1.4cm}{\centering average initial haplotypes} }&
\multicolumn{2}{c}{\parbox{1.4cm}{\centering average result size} }&
\multicolumn{2}{c}{\parbox{1.4cm}{\centering average ratio\\$r$} } \\
\midrule
50      & 50    & 0     & \hspace{4.3mm}18&.5 & \hspace{4.3mm}19&.5 & \hspace{4.3mm}19&.5 & \hspace{5mm}1&   \\% & 8&.66 \\
        &       & 8     & 20&.13        & 21&.73        & 22&.03        & 1&.02\\% & 8&.33 \\
        &       & 16    & 22&.27        & 24&.77        & 25&.77        & 1&.04\\% & 9&.04 \\
        &       & 24    & 20&.63        & 24&.17        & 25&.23        & 1&.05\\% & 10&.38 \\
\cmidrule(lr){2-11}
        & 75    & 0     & 22&.1 & 23&.13        & 23&.1 & 1&   \\% & 7 \\
        &       & 8     & 24&.63        & 26 &  & 26&.27        & 1&.01\\% & 7&.06 \\
        &       & 16    & 25&.6 & 27&.3 & 27&.37        & 1&.01\\% & 7&.42 \\
        &       & 24    & 25&.3 & 27&.63        & 28&.1 & 1&.02\\% & 8&.47 \\
\cmidrule(lr){2-11}
        & 100   & 0     & 25&.07        & 26&.13        & 26&.07        & 1&   \\% & 6&.69 \\
        &       & 8     & 26&.7 & 27&.93        & 27&.8 & 1&   \\% & 6&.49 \\
        &       & 16    & 28&.27        & 29&.87        & 29&.73        & 1&   \\% & 6&.64 \\
        &       & 24    & 28&.5 & 30&.5 & 30&.2 & 0&.99\\% & 7&.1 \\
\midrule
75      & 50    & 0     & 21&.77        & 22&.77        & 22&.77        & 1&   \\% & 16&.37 \\
        &       & 8     & 23&.1 & 25&.37        & 26&.63        & 1&.05\\% & 19&.77 \\
        &       & 16    & 24&.97        & 29&.77        & 34&.4 & 1&.16\\% & 25&.68 \\
        &       & 24    & 25&.1 & 31&.4 & 38&.33        & 1&.23\\% & 29&.39 \\
\cmidrule(lr){2-11}
        & 75    & 0     & 26&.17        & 27&.2 & 27&.17        & 1&   \\% & 16&.13 \\
        &       & 8     & 29&.9 & 31&.5 & 31&.77        & 1&.01\\% & 17&.27 \\
        &       & 16    & 30&.93        & 34&.63        & 37&.1 & 1&.07\\% & 20&.99 \\
        &       & 24    & 31&.23        & 35&.83        & 38&.83        & 1&.08\\% & 25&.03 \\
\cmidrule(lr){2-11}
        & 100   & 0     & 29&.5 & 30&.5 & 30&.5 & 1&   \\% & 17&.27 \\
        &       & 8     & 32&.87        & 34&.23        & 34&.13        & 1&   \\% & 16&.19 \\
        &       & 16    & 33&.67        & 36&.1 & 36&.77        & 1&.02\\% & 18&.99 \\
        &       & 24    & 36&.2 & 39&.73        & 41&.17        & 1&.04\\% & 20&.84 \\
\midrule
100     & 50    & 0     & 24&.33        & 25&.33        & 25&.33        & 1&   \\% & 32&.16 \\
        &       & 8     & 27&   & 30&.67        & 36&.6 & 1&.2 \\% & 47&.17 \\
        &       & 16    & 27&.53        & 32&.8 & 41&.8 & 1&.28\\% & 51&.13 \\
        &       & 24    & 27&.93        & 36&.1 & 49&   & 1&.36\\% & 65&.34 \\
\cmidrule(lr){2-11}
        & 75    & 0     & 27&.83        & 28&.83        & 28&.83        & 1&   \\% & 31&.62 \\
        &       & 8     & 32&.2 & 34&.4 & 36&.07        & 1&.05\\% & 35&.74 \\
        &       & 16    & 34&.23        & 38&.33        & 43&   & 1&.12\\% & 39&.82 \\
        &       & 24    & 35&.23        & 42&.07        & 50&.43        & 1&.2 \\% & 48&.15 \\
\cmidrule(lr){2-11}
        & 100   & 0     & 34&.5 & 35&.5 & 35&.5 & 1&   \\% & 31&.37 \\
        &       & 8     & 37&.37        & 39&.13        & 40&   & 1&.02\\% & 33&.57 \\
        &       & 16    & 39&.87        & 43&.27        & 45&.63        & 1&.06\\% & 37&.3 \\
        &       & 24    & 38&.6 & 45&.13        & 52&.87        & 1&.17\\% & 46&.68 \\
\bottomrule
\end{tabular}
\end{table*}

%\notaestesa{YP}{COMMENTO AI RISULTATI QUI}
On both datasets the heuristic produces comparable results.
In particular, the average ratio is never larger than $1.57$, while quite
often it is close to $1$.
In other words, it can often reconstruct a solution of size similar to
the number of the haplotypes used to generate the instance and, in the
worst case, the computed solution is at most $1.57$ larger than the set
of initial haplotypes.
The ability of computing a good approximation seems affected by two
combined factors: the number of independent characters of the genotype
matrix and the number of initial haplotypes.
Indeed in both tables we can observe that the smaller the number of
independent characters compared to the number of initial haplotypes, the
worse is the computed solution.
Conversely good solutions are computed by the heuristic when the number
of independent characters is close to the number of initial haplotypes.

Lemma~\ref{lem:lb-m} offers a possible explanation to such
regular behavior of our heuristic.
In fact, let $H$ be the set of initial haplotypes of an instance $X$ and
suppose that they are defined on a set $\Sigma$ of independent
characters such that $\card{H} = \card{\Sigma} + 1$ (i.e.~$H$ is also a
solution that meets the lower bound of Lemma~\ref{lem:lb-m}).
Then, the set $T$ computed during step~\ref{alg:heu:gauss} of the
heuristic algorithm contains exactly $\card{\Sigma}$ independent
xor-genotypes.
As a consequence, the set $C$ computed in the same
step admits a Graph Realization and, thus, the heuristic solves
optimally the instance $X$.
Although this is not the general case, our intuition suggests that, when
the number of independent characters is close to the number of initial
haplotypes, the selection of the set $T$ is constrained and the Graph
Realization of the maximal subset of $C$ computed by the heuristic is
similar to the xor-graph
associated with the initial haplotypes.
Conversely, if the number of independent characters is significantly
lower than the number of initial haplotypes, there are a lot of degrees
of freedom in the choice of the set $T$, thus the output of the Graph
Realization step can vary greatly from the
xor-graph of the initial haplotypes.

The time required by the heuristic to compute a solution to
the pure-random synthetic instances varies between circa 25 seconds on
instances with 100 genotypes and 70 minutes on instances with 400
genotypes.
All the instances generated using the neutral model, instead, have been
solved in less than 1 minute.
We also observe that instances where the heuristic fails to find a good
solution have been solved considerably faster than the ones where the
heuristic computes a good approximation.
However, a more careful analysis suggests that such fluctuations are due
to the different amount of I/O operations needed to communicate with the
\texttt{GREAL} software that we use to solve the Graph Realization
problem.

Finally we tried to compare our heuristic method with the
ILP formulation proposed by Brown and Harrower~\cite{BHAR}.
In the paper, they formulate the PPXH problem as a polynomial-size
integer linear program and they introduce cuts and modification of the
objective function that should help finding the optimal solution.
However, the \texttt{GLPK} solver~\cite{GLPK}, using the basic
formulation as well as the augmented formulations, was not able to find
a feasible solution even for the smallest instances of our
experimentation (50 genotypes and 50 characters) within the maximum time
of 24 hours.

\subsubsection{Real Data}
To validate the feasibility of applying our heuristic on
real data, we have produced some instances from the Phase I
dataset of the HapMap
project \cite{haplomap} (release 2005-06\_16c.1).
A set of xor-genotypes were produced from the data for each
population in the dataset (discarding non biallelic sites and non
autosomal chromosomes).
Those instances vary from  44 genotypes and
184604 sites to 90 genotypes and 91812 sites.
On average, an instance contains 67 genotypes and 46906 sites.
On all those instances our heuristics has never required more than
$2$ seconds on the same PC used in the experimental part over synthetic instances,
clearly establishing that the heuristic can be successfully used
on real-world large instances.

\section{Conclusions and Future Work}
% of course!!

In the paper we investigate the problem of resolving xor-genotypes
under the pure parsimony model. We give several results regarding
the efficient solution of the problem by considering fixed-parameter
algorithms  or by  restricting  the instances of the
problem. Most of the results are based on combinatorial
properties of a graph representation relating a feasible
solution to the instance: the
xor-graph. The computational complexity of the unrestricted problem is
still unknown. Since we show that PPXH($\infty,2$) and PPXH($2,
\infty $) have  polynomial time algorithms, it would be interesting to
determine the complexity of PPXH($\infty,3$)  and PPXH($3, \infty
$), as these two cases could delimit
polynomial time solvability and intractability of the general
problem.  We believe that the xor-graph could play a crucial role
in solving these open problems.

  \section*{Acknowledgments}

PB, GDV and YP have been partially supported by FAR 2008 grant
  ``Computational models for phylogenetic analysis of gene variations''.
PB has been partially supported by
the MIUR PRIN 2007 Project  ``Mathematical aspects and emerging
applications of automata and formal languages''.

\bibliographystyle{abbrv}
\bibliography{abbreviations,books,biology,complexity,splicing,haplotype,sw,graphs,general-algo}

\end{document}